\documentclass[twocolumn,aps,prl,superscriptaddress]{revtex4}

\usepackage{amsmath}
\usepackage{graphicx}
\usepackage{MnSymbol}

\begin{document}

%\title{Sphere and cylinder contact mechanics during slip}
\title{On the origin of sliding friction: Role of lattice trapping}

\author{J. Wang}
\affiliation{PGI-1, FZ J\"ulich, Germany, EU}
\affiliation{College of Science, Zhongyuan University of Technology, Zhengzhou 450007, China}
\author{ A. Tiwari}
\affiliation{PGI-1, FZ J\"ulich, Germany, EU}
\affiliation{www.MultiscaleConsulting.com}
\author{ I.M. Sivebaek}
\affiliation{PGI-1, FZ J\"ulich, Germany, EU}
\affiliation{Department of Mechanical Engineering, Technical University of Denmark, Lyngby 2800, Denmark}
\affiliation{Novo Nordisk Device R\&D, 3400 Hiller$\phi$d, Denmark}
\author{ B.N.J. Persson}
\affiliation{PGI-1, FZ J\"ulich, Germany, EU}
\affiliation{www.MultiscaleConsulting.com}

%\author[1, 2]{\small J. Wang}
%\author[1, 3]{\small A. Tiwari}
%\author[1, 3, 4]{\small I. Sivebaek}
%\author[1, 3]{\small B. N. J. Persson}
%
%\affil[1]{\footnotesize PGI-1, FZ J\"ulich, Germany, EU }
%\affil[2]{\footnotesize College of Science, Zhongyuan University of Technology, Zhengzhou 450007, China}
%\affil[3]{\footnotesize www.MultiscaleConsulting.com}
%\affil[4]{\footnotesize Department of Mechanical Engineering, Technical University of Denmark, Produktionstorvet, Building 427, Kongens Lyngby 2800, Denmark}
%\affil[5]{\footnotesize Novo Nordisk Device R \& D, DK-400 Hiller$\phi$d, Denmark}

\begin{abstract}
Using molecular dynamics we study the dependence of the friction force on the sliding speed when an elastic
slab (block) is sliding on a rigid substrate with a ${\rm sin} (q_0 x)$ surface height profile. 
The friction force is nearly velocity independent due to phonon emission at the closing and opening crack tips, where rapid atomic snap-in and -out events occur during sliding. The rapid events result 
from lattice trapping and are closely related to
the velocity gap and hysteresis effects observed in model studies of crack propagation in solids.
This indicates that the friction force is dominated by processes 
occurring at the edges of the contact area, which is confirmed 
by calculations showing that the friction force is independent of the normal force.
The friction force increases drastically when the sliding velocity approaches the solid transverse sound velocity, as expected from the theory of cracks.
\end{abstract}

\maketitle
%\makenomenclature

%%%%%%%%%%%%%% main text %%%%%%%%%%%%%%%%
%\begin{multicols}{2}

{\it Introduction}--The friction force acting on a block sliding on a substrate
is usually nearly velocity independent unless the sliding
speed is so low that thermal activation is important, or so high that frictional heating becomes important.
A velocity independent friction force results if rapid processes occur at the sliding interface, involving
local slip velocities unrelated to the macroscopic drive velocity. One important topic in tribology is to understand the origin and nature of the rapid slip events, which generate the sliding friction force\cite{P3}.

The friction force is usually proportional to the normal force (Amonton's law). This result follows quite
general from the theory of the contact between elastic solids with random surface roughness\cite{Am,super}.
Thus, the area of real contact is usually proportional to the
normal force\cite{Hyun,JCP,SSR,Mueser,Prod,Camp,Carb}. This results from the fact that for a large system, when the normal force increases 
the number of asperity contact region increases proportional to the normal force,
but the distribution of asperity contact areas and the (contact) stress distribution are unchanged\cite{add1,add2}. 
It follows that
the friction force will be proportional to the normal force independent of the nature of the microscopic
frictional interaction at the asperity level, e.g., independent of how the friction force acting on an asperity 
depends on the asperity contact area.

\begin{figure}
\centering
\includegraphics[width=0.4\textwidth]{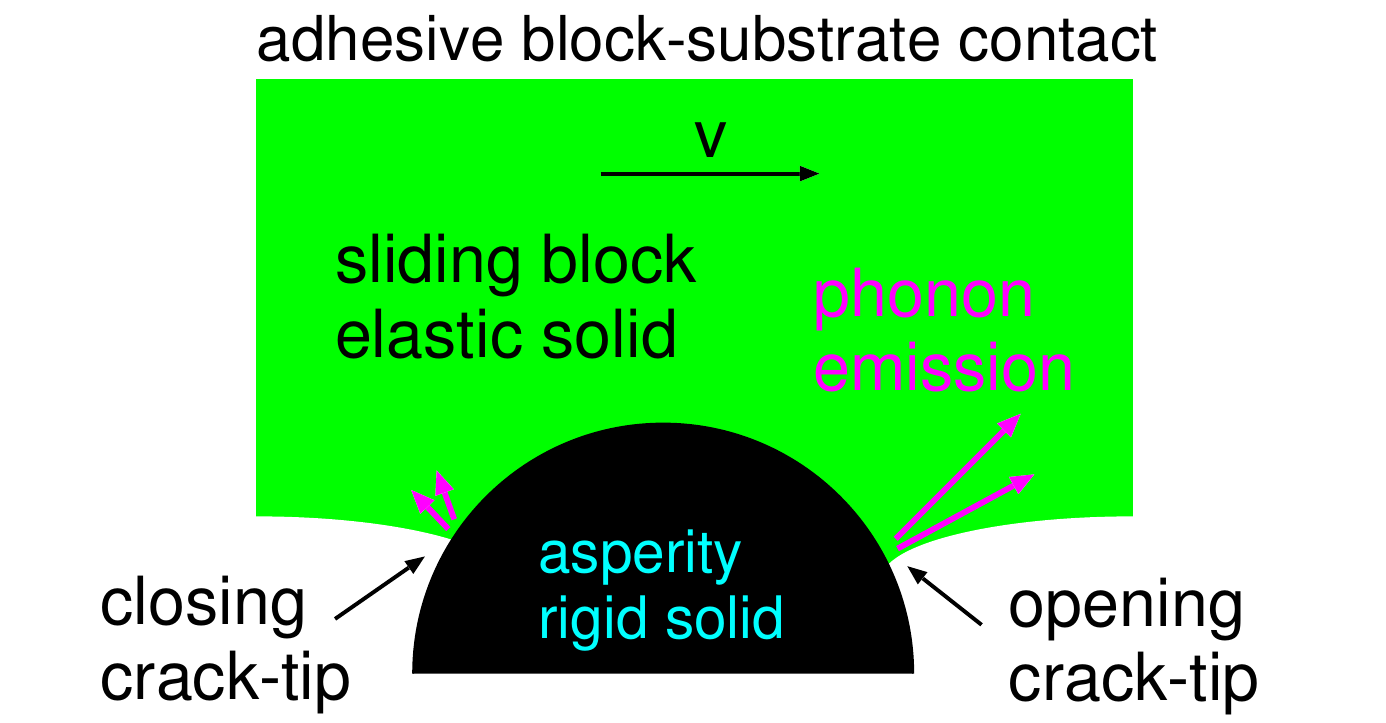}
\caption{\label{AsperityContact} 
An elastic block (green) sliding on a substrate (black). At the opening and closing crack tips
rapid atomic snap-off and snap-in processes occur which is the origin of the observed sliding friction.
}
\end{figure}

In this paper, we present a molecular dynamics (MD) study of the dependency
of the friction force on the sliding velocity when an elastic
slab (block) is sliding on a rigid substrate with a ${\rm sin} (q_0 x)$ surface height profile. The atoms on the
block interact with the substrate atoms by Lennard-Jones potentials. 
We show that the friction force 
is due to lattice pinning: at the opening and (to lesser extent) the closing crack tips 
atoms snap-out (and -in) of contact in rapid events, with atom velocities unrelated to the block driving speed,
followed by ``long'' time periods where the crack tips are pinned. In the rapid
slip events elastic waves (phonons) are emitted from the crack tips (see Fig. \ref{AsperityContact}), resulting 
(for the opening crack) in a larger crack propagation energy than the
adiabatic value. 

\begin{figure}
\centering
\includegraphics[width=0.45\textwidth]{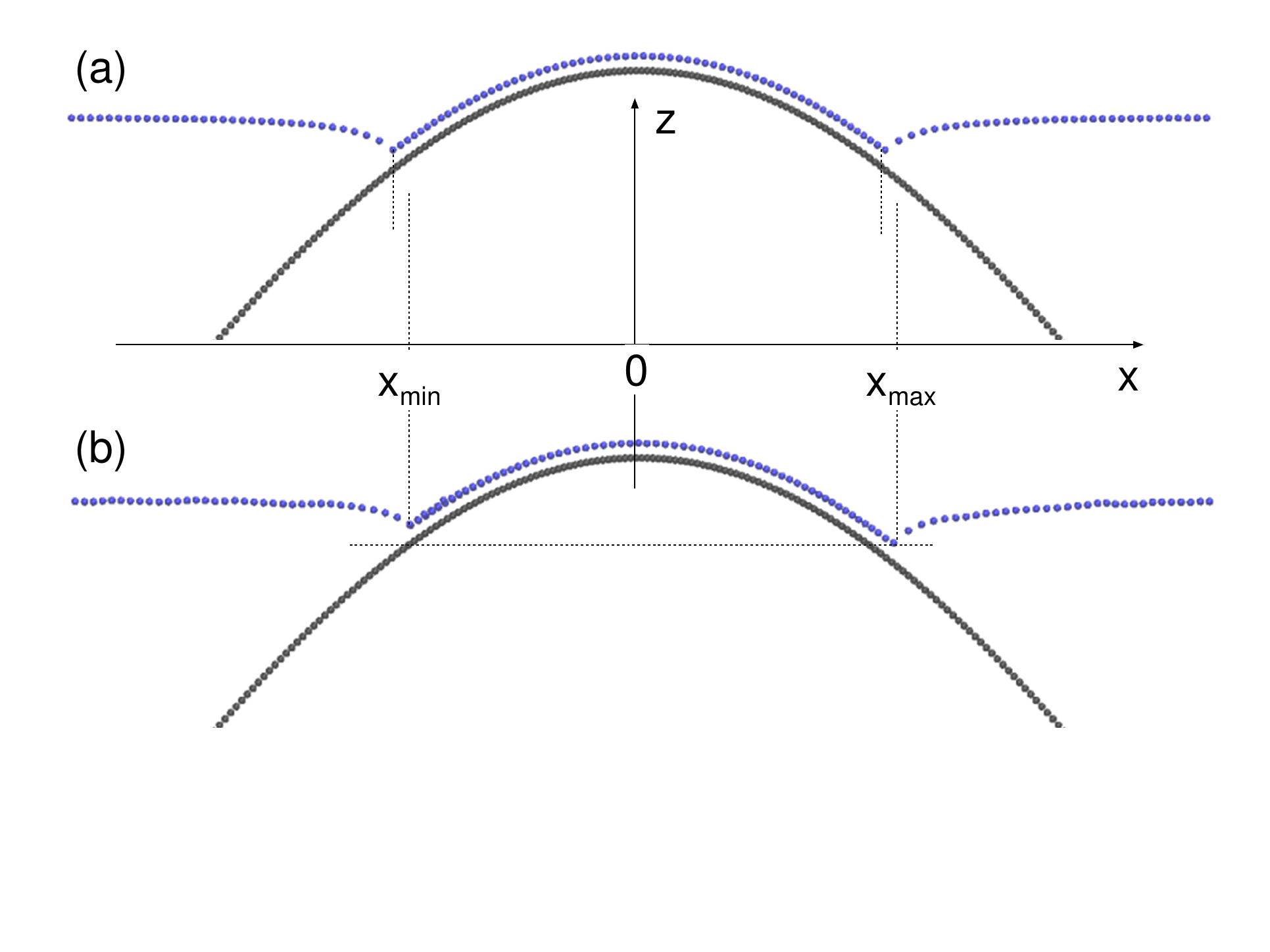}
\caption{\label{ContactPic} The contact area between an elastic slab (block) and a rigid substrate at the temperature $T=0 \ {\rm K}$.
The substrate is corrugated with the height coordinate $z=h_0 {\rm sin} (q_0 x)$ ($h_0 = 100  \ {\rm \AA}$ and $q_0 =  \pi /L_x$ with $L_x=254 \ {\rm \AA}$). The nominal contact pressure $p= F_z /(L_x L_y) = 0.1 \ {\rm MPa}$. (a) before start of sliding and (b) after sliding
$3 \ {\rm nm}$ at the velocity $v=0.1 \ {\rm m/s}$.}
\end{figure}

\vskip 0.1cm
{\it Model}--We consider the contact between an elastic slab and a rigid substrate with the cylinder corrugation (see Fig. \ref{ContactPic})
$z=h_0 {\rm sin}(q_0 x)$,
where $q_0 = \pi /L_x$ and $0<x<L_x$. We assume periodic boundary conditions in the $xy$ plane with the basic unit having the dimensions
$L_x= 254 \ {\rm \AA}$ and $L_y=14 \ {\rm \AA}$. The corrugation amplitude $h_0 = 100 \ {\rm \AA}$.
In order for the contact mechanics not to depend on the block thickness, one must choose
the block thickness larger than the diameter of the block-substrate contact region. In the present study, the block thickness is
$d \approx 276  \ {\rm \AA}$.

The substrate is rigid.
The springs between the block atoms have
elongation and bending stiffness so chosen as to reproduce Young's modulus $E$ and shear modulus
$G$ specified as input for the calculations, and here we use $E=10 \ {\rm MPa}$
and $G=3.33 \ {\rm MPa}$, corresponding to the Poisson ratio $\nu \approx 0.5$.

Fig. \ref{ContactPic}(a) 
shows the contact between the elastic slab (block) and the substrate at the
temperature $T=0 \ {\rm K}$ before start of sliding. We only show the first layer of atoms of the block and the substrate at the interface.
The substrate and the block have $N_x=206$ and $128$ atoms along a row in the $x$-direction, and $N_y=11$ and $7$ atoms in the $y$-direction,
respectively.
The substrate and block-lattice constants $a_{\rm s} = L_x/N_x \approx 1.233 \ {\rm \AA}$
and $a_{\rm b} \approx 1.984 \ {\rm \AA}$, respectively. 
The ratio $a_{\rm b}/a_{\rm s} \approx 1.609$ is close to the golden mean $(1+\surd 5)/2\approx 1.618$ so the contact is ``almost'' incommensurate.
The block mass density $\rho = m/a_{\rm b}^3 = 1060 \ {\rm kg/m^3}$.

The atoms at the interface between the block and the substrate interact via
the Lennard-Jones (LJ) interaction potential:
$$V(r) = 4 V_0 \left [\left ({r_0\over r}\right )^{12}-\left ({r_0\over r}\right )^{6}\right ],$$
where $V_0 = 0.04 \ {\rm eV}$ and $r_0=3.28 \ {\rm \AA}$.
This LJ potential gives the adiabatic work of adhesion $w_0 \approx 0.0027 \ {\rm J/m^2}$.
This work of adhesion is very small, and a more typical value is $w_0 \approx 0.1 \ {\rm J/m^2}$, but this
would in the present case result in complete contact at the interface.
If the attractive part of the LJ potential is removed (i.e., no adhesion)
a ``superlubric'' sliding state would prevail, with vanishing sliding friction.

During sliding lattice vibrations (phonons)
are emitted from the contact region, and for a finite system
without internal damping, the block will heat up and after long enough sliding distance the thermal fluctuations will influence the
contact mechanics and the friction force.
For this reason, it is important to choose the thickness of the sliding block relatively large.
The thicker this layer is the smaller influence will the thermal fluctuations,
resulting from emitted phonons, have on the contact mechanics.

In the present study, we include a Langevin type of damping force (proportional to the atom relative velocity)
in the equation of motion for the block atoms during the initial contact formation (no sliding).
After we have obtained the initial contact state (at zero temperature) we remove the damping term and consider so short sliding distances
that frictional heating is negligible.

In Fig.  \ref{ContactPic}(a)
we show pictures of the contact after squeezing the solids into contact (no sliding) 
for the nominal contact pressure $p= F_z /(L_x L_y) = 0.1 \ {\rm MPa}$. 
Fig.  \ref{ContactPic}(b) shows the contact after 3 nm of sliding at 0.1 m/s. 

\begin{figure}
\centering
\includegraphics[width=0.45\textwidth]{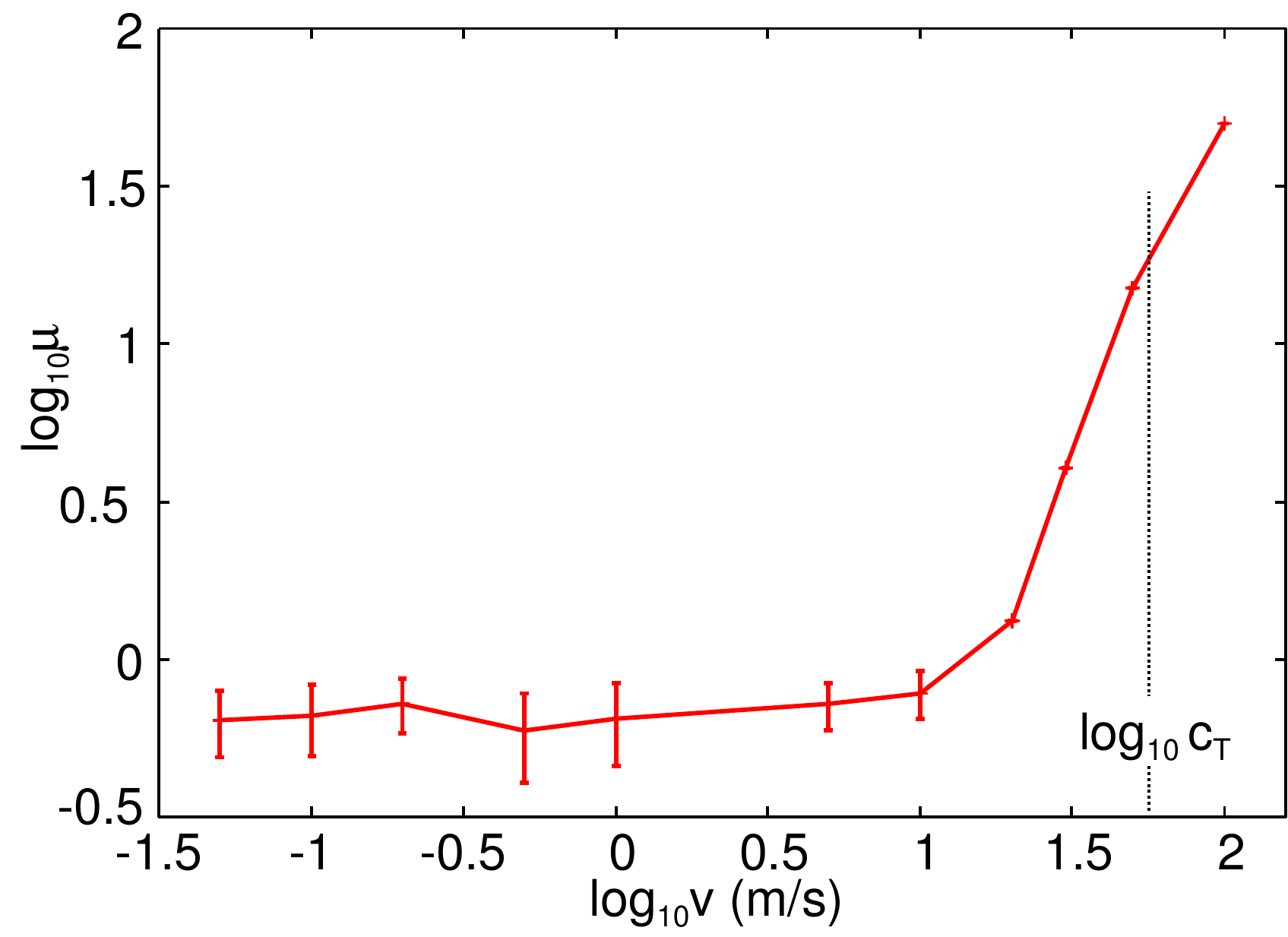}
\caption{\label{friction}
The friction coefficient $\mu = F_x/F_z$ as a function of the logarithm of the sliding speed and for the nominal contact pressure
$p=0.1 \ {\rm MPa}$. The vertical dashed line is for $v=c_{\rm T}$, where the transverse sound velocity $c_{\rm T}=56 \ {\rm m/s}$.}
\end{figure}

\begin{figure}
\centering
\includegraphics[width=0.45\textwidth]{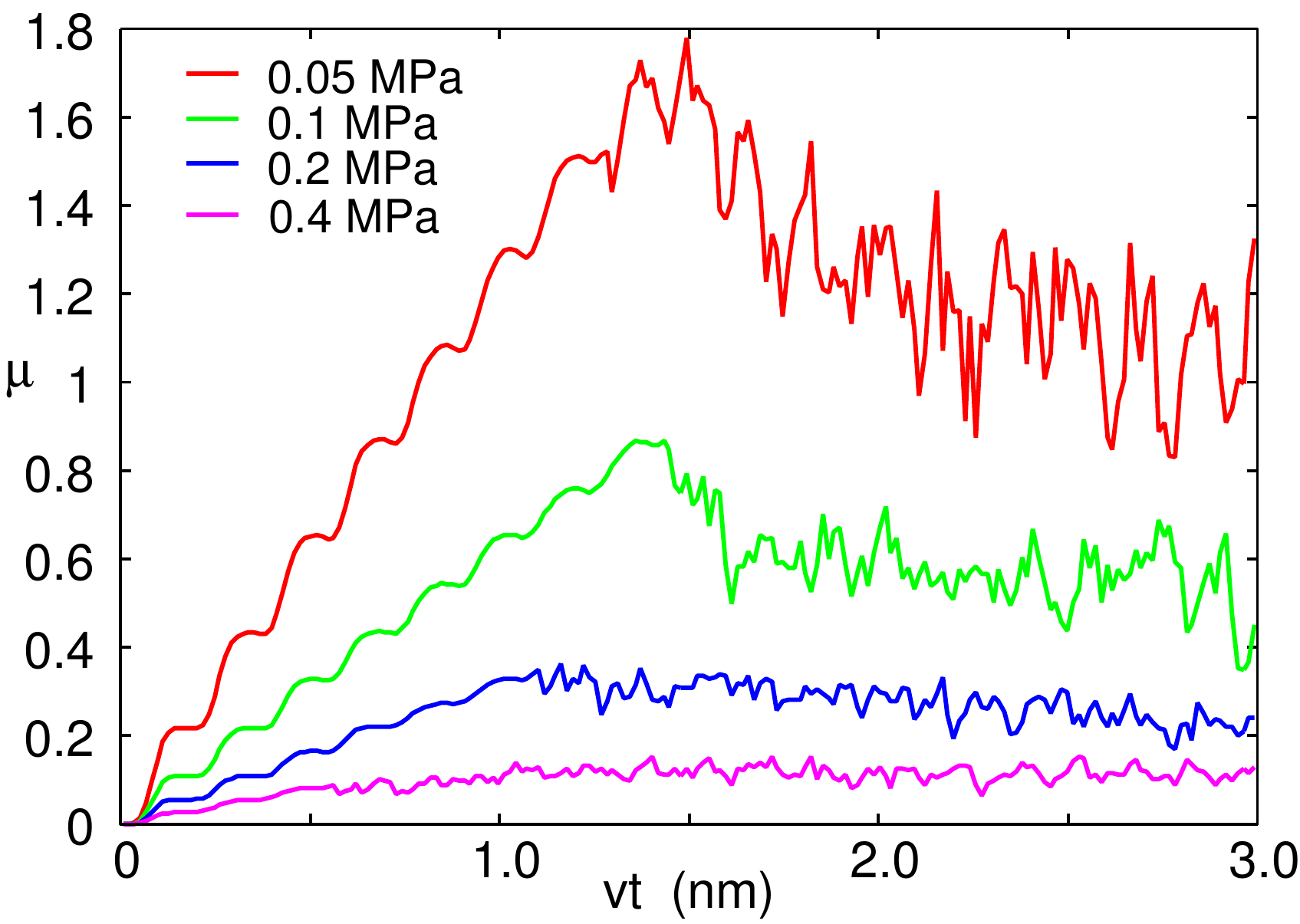}
\caption{\label{test-pressure-cof.pdf}
The friction coefficient as a function of the sliding distance and the nominal contact pressure
$p=F_z/(L_xL_y)$. For the case of adhesion and the sliding speed $v=0.1 \ {\rm m/s}$.
}
\end{figure}

\begin{figure}
\centering
\includegraphics[width=0.5\textwidth]{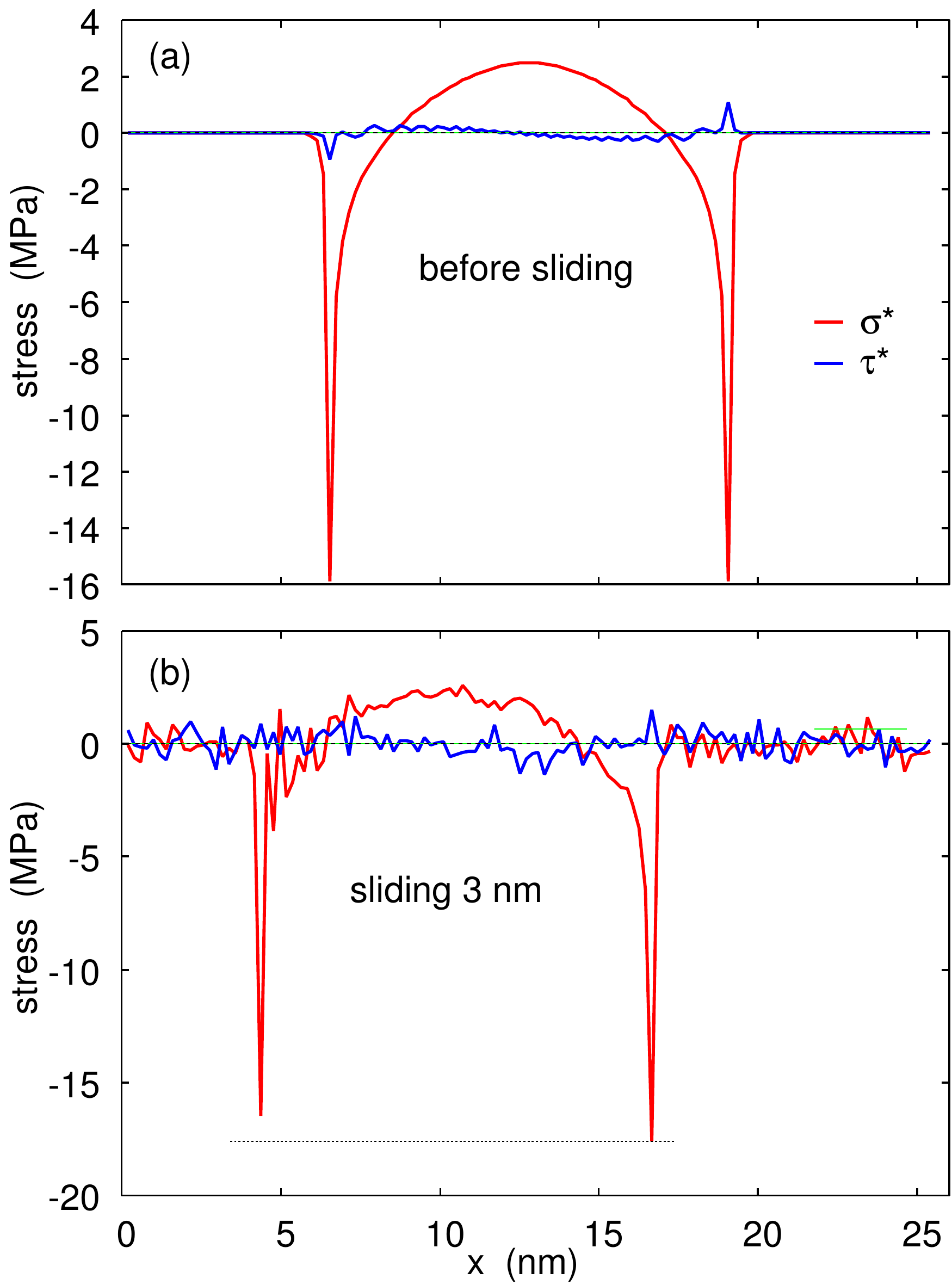}
\caption{\label{1x2sigmaStaadhesion}
The normal stress $\sigma^*$  (red line) and the shear stress $\tau^*$ (blue)
acting on the block as a function of the spatial coordinate $x$.
(a) Only squeezing and (b) after sliding $3 \ {\rm nm}$ at the sliding speed $v=0.1 \ {\rm m/s}$.
The nominal contact pressure $p=0.1 \ {\rm MPa}$.
}
\end{figure}

\vskip 0.1cm
{\it Results}--Fig. \ref{friction} shows the  logarithm of the kinetic friction coefficient $\mu = F_x/F_z$ as a
function of the logarithm of the sliding speed.
The friction coefficients were obtained after sliding $3 \ {\rm nm}$ at the given sliding speeds.
The nominal contact pressure acting on the upper surface of the block is $p=0.1 \ {\rm MPa}$.
Note that for $v < 10 \ {\rm m/s}$ the friction coefficient
is nearly velocity independent and equal to $\approx 0.6$. 

The near velocity independence of the friction force is due to the fact that the
friction is caused by rapid slip events, where the local slip velocity is unrelated to
the driving speed. The local slip events are easily observed at the opening crack tip where atoms jump (snap)
out of contact in very rapid events, followed by ``long'' time periods where the tip
is pinned by the corrugated interfacial atomic interaction potential (see movies online in Ref. \cite{movie}).
During the rapid snap-out of
contact elastic waves (phonons) are emitted from the opening crack tip,
and this is the origin of the friction force\cite{Ho, commentHu}.
This effect is closely related to lattice trapping, the velocity gap
and hysteresis effects observed in model studies of crack propagation in solids\cite{Marder,Per1,Per2,Per3,Per4}.

If $F_{\rm f}$ denote the friction force, the dissipated energy during sliding of the distance $L$ is $F_{\rm f} L$, 
and if the friction is due entirely to
energy dissipation at the opening and closing crack tips, then $F_{\rm f} L = (w_{\rm open}-w_{\rm close})LL_y$,
where $w_{\rm open}>w_0$ and $w_{\rm close}<w_0$ are the opening and closing 
crack propagation energies (per unit surface area). The friction
coefficient $\mu = F_{\rm f}/F_z$ with the normal force $F_z = L_x L_y p$. Thus 
$$\mu= {w_{\rm open}-w_{\rm close}\over pL_x}. \eqno(1)$$
Using $L_x =254 \ {\rm \AA}$, $p=0.1 \ {\rm MPa}$
and $\mu \approx 0.6$ this gives $w_{\rm open}-w_{\rm close} \approx 0.0015 \ {\rm J/m^2}$. The crack propagation hysteresis factor 
$Q=(w_{\rm open}-w_{\rm close})/w_0 \approx 0.56$ is very similar to the hysteresis (due to lattice trapping)
observed in atomistic MD crack propagation calculations\cite{Marder,Per1,Per2,Per3,Per4}, e.g., $Q \approx 0.45$ for zero temperature for 
the 1D-string model studied in Ref. \cite{Per2}.

Since $w_{\rm open}$ and $w_{\rm close}$ are independent of the applied pressure $p$, 
(1) predicts that the friction coefficient $\mu \sim 1/p$, i.e., the friction force is independent of the applied normal force. 
To test this we have performed MD simulations 
with $p=0.05$, $0.1$, $0.2$ and $0.4 \ {\rm MPa}$, see Fig. \ref{test-pressure-cof.pdf}. 
Clearly, within the noise of the calculations, the friction coefficient is proportional to
$1/p$ confirming that the friction is entirely due to the emission of phonons from the opening and the closing
crack tips.

The calculation above is for $T=0 \ {\rm K}$. As the temperature increases, the hysteresis factor $Q(T)$ decreases\cite{Per2}.
Hence the contribution to the sliding friction from the phonon emission from the crack tips decreases with increasing temperature.

From the continuum mechanics theory of cracks, it is known that the crack propagation energy (per unit created surface area) diverges when
the crack tip velocity approaches the velocity of elastic wave propagation in the solids 
(more exactly, the Rayleigh sound speed)\cite{div}. This is due to the emission of elastic waves (phonons) from the moving crack tip.
In the present case the transverse sound velocity $c_{\rm T} = (G/\rho)^{1/2} \approx 56 \ {\rm m/s}$ (the Rayleigh sound speed
$c_{\rm R} \approx 0.95 c_{\rm T}$). Hence we expect the friction to increase drastically as the drive velocity $v$ approaches $c_{\rm T}$,
as it is indeed the case (see Fig. \ref{friction}).

The emission of sound waves from the opening crack results in a crack propagation energy which is
larger than the adiabatic value, while for the closing crack it is smaller than the adiabatic value.
This results in an asymmetric contact where $x_{\rm max} > |x_{\rm min}|$.
This asymmetry is easily observed in pictures of the interfacial separation as a function of the
lateral coordinate $x$; see Fig. \ref{ContactPic}(b) and Ref. \cite{Nano}.

Let us now study the stresses acting normal and tangential to the (rigid) substrate profile.
These stresses, which we denote as $\sigma^*$ and $\tau^*$, respectively,
can be easily obtained from the linear combination of $\sigma=\sigma_{zz}$ and $\tau=\sigma_{xz}$:
$\sigma^* = \sigma {\rm cos}\theta -\tau {\rm sin}\theta$ and $\tau^*  = \sigma {\rm sin}\theta +\tau {\rm cos}\theta$,
where ${\rm tan}\theta = z'(x)=q_0h_0 {\rm cos} (q_0 x)$ is the slope of the substrate profile.
In Fig. \ref{1x2sigmaStaadhesion} we show  $\sigma^*$ and $\tau^*$
as a function of the spatial coordinate $x$.
The nominal contact pressure $p=0.1 \ {\rm MPa}$ and the sliding speed
$v=0.1 \ {\rm m/s}$. We show results (a) after squeezing the solids into contact (zero sliding distance),
and (b) after sliding $3 \ {\rm nm}$. Note the large adhesive stress at the edges (crack tips) of the contact region. If $r$ denotes the
distance from a crack tip, from the continuum model
of adhesion (the JKR theory for stationary contact), one expects the stress 
to diverge as $r^{-1/2}$ as one approach a crack tip\cite{div}.

Note that the friction force is the integral over the surface area, from $x=0$ to $x=L_x$,
of the stress $\sigma_{xz} = \tau^* {\rm cos} \theta-\sigma^* {\rm sin}\theta$.
The shear stress $\tau^*$ in the contact area takes both positive and negative values
and the integral of $\tau^* {\rm cos} \theta$ over $x$ nearly vanish. Hence, the biggest contribution to the friction force comes from the
stress $\sigma^*$ normal to the substrate profile. For the sliding state (Fig. \ref{1x2sigmaStaadhesion}(b))
the integral of $\sigma^* {\rm sin}\theta$ over $x$ is nonzero 
and relative large because
the adhesive stress $\sigma^*$ at the opening crack tip is considerably larger than the
corresponding stress at the closing crack tip (see Fig. \ref{1x2sigmaStaadhesion}(b)).

%In our simulations we observe a non-uniform shear stress. This is due to the fact that we use a very small work of adhesion
%$w=0.0027 \ {\rm J/m^2}$, so that $w_0/d \approx 2 \ {\rm MPa}$ while the contact pressure
%reaches $\approx 10 \ {\rm MPa}$ at the opening and closing
%crack tips (see Fig. \ref{1x2sigadhesion}). In addition, inside the contact the atomic positions of the
%block are nearly incommensurate with respect to
%the substrate atoms so the main frictional interactions occur at the opening and closing crack tips.
%This differs from the case of rubber on a glass surface where
%the rubber chains at the interface rearrange themselves in the substrate potential and pin the surfaces together.
%During sliding nanometer sized pinned regions
%undergo stick-slip type of motion, resulting in a frictional shear stress which is nearly uniform in the contact region.

\begin{figure}
\includegraphics[width=0.95\columnwidth]{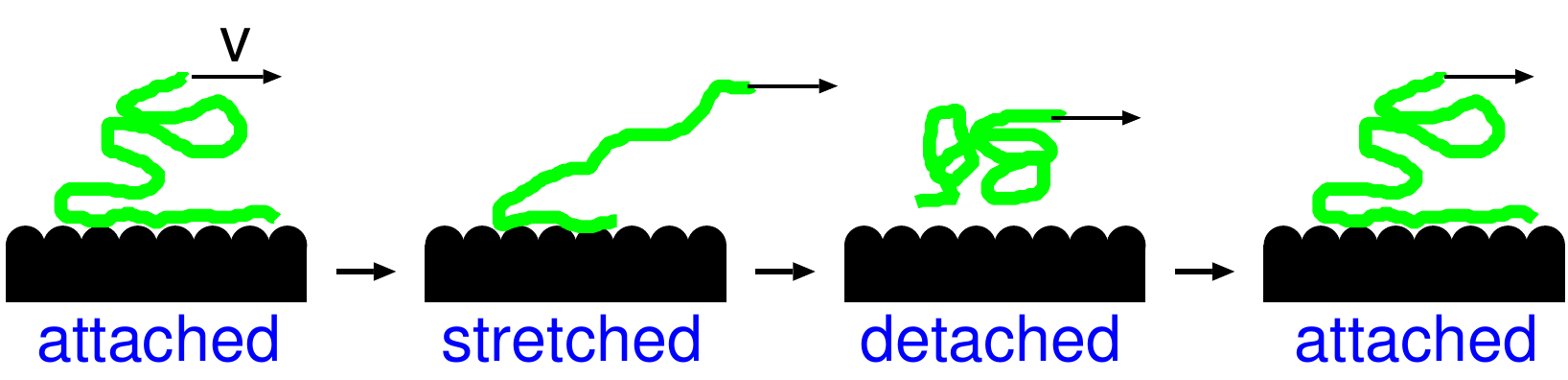}
\caption{\label{PerssonVol}
The classical description of a polymer chain at the rubber surface. During the lateral
motion of the rubber block the chain stretches, detaches, relaxes, and re-attaches to the substrate surface to repeat the cycle.
The picture is schematic and in reality no complete detachment in the vertical direction is expected, but only a rearrangement
of molecule segments (in nanometer-sized domains) parallel to the surface from pinned (commensurate-like) to depinned
(incommensurate-like) domains.
}
\end{figure}

\vskip 0.1cm
{\it Discussion}--We have shown that the friction force is nearly velocity independent for
$v<<c_{\rm T}$, where $c_{\rm T}$ is the velocity of transverse sound waves in the block.
The friction force is mainly due to energy
dissipation at the opening crack tip, where rapid atomic snap-off events occur
during sliding. This ``edge-dominated friction'' is very different from the frictional
processes found when a macroscopic silicone rubber sphere is sliding on a substrate.
In the latter case, one observes an ``area-dominated friction'' where the shear stress is nearly uniform
within the contact area\cite{Chat,Ion}. In this case, the  friction force
arises from the stick-slip type of motion of nanometer-sized regions 
everywhere within the contact region (see Fig. \ref{PerssonVol})\cite{Schall,smooth}.

The difference between the system we study (crystalline elastic block) 
and (silicone) rubber is that rubber materials have nanometer-thin surface
layers where the polymer chains have large (liquid-like) mobility.  
In this case, the rubber chains at the interface rearrange themselves in the substrate potential 
forming commensurate-like (nanometer-sized) domains, which pin the surfaces together. During sliding pinned regions
undergo stick-slip type of motion, resulting in frictional shear stress which 
is nearly uniform in the contact region. In the case, we studied above both the substrate and the
block are crystalline materials with (nearly) incommensurate structures, and in this case there is a negligible contribution
to the friction from the internal area of the contact region. We note, however,
for larger contact areas, with stronger adhesive interaction and for elastically softer materials,
we expect the formation of nano-sized commensurate-like regions (stress domains\cite{stress}) 
at the contact interface, and in this case we expect
a contribution to the friction also from the internal regions of the contact. 

The phonon-emission processes associated with the opening and closing crack tips
are likely to be insensitive to the linear size of the block-substrate contact region. For viscoelastic solids
like rubber, there is a viscoelastic contribution to the crack-opening energy, which may involve regions in the solid
far away from the crack tip, which for high enough crack tip speed may enhance the crack propagation energy with a very large factor,
given by the ratio between Young's modulus in the glassy and the 
rubbery region (enhancement factor typically of order 100-1000)\cite{Gennes,Brener,Gert,small}. 
Thus, for high enough sliding speeds the crack
tip region may give a very important contribution to the sliding friction 
force for viscoelastic solids. We note, however,
that the higher the sliding speed the further away from the crack tip the dominating 
viscoelastic energy dissipation will occur, which will result in a finite-size effect: if the
asperity contact region is small the viscoelastic contribution to the crack propagation energy 
may be strongly reduced\cite{small}.

Most real surfaces have layers of weakly adsorbed molecules, e.g., hydrocarbons. 
In this case too, one expects an important contribution to the friction force from the internal area of the contact region.
Thus, when weakly bound ``contamination'' molecules are located between two solids 
they will adjust to the corrugated potential of both walls and pin the surfaces together.
This will result in a non-zero breakloose (or static) friction force.  
During sliding instabilities occur where the molecules 
rapidly slip at velocities unrelated to the (macroscopic) block driving speed. 
After each slip event, the local vibrational motion may occur, which is damped by 
phonon emission\cite{Ho,commentHu}, very similar to the processes occurring at the opening crack tip in the 
model studied above. At low temperature this usually results in a kinetic friction force which is
nearly independent of the sliding speed, except at very low sliding speeds where thermal activation becomes important,
where the friction force depends logarithmic or linearly on the sliding speed\cite{P3,He,Mus}.

Finally, we note that elastically hard materials like diamond usually exhibit very low sliding friction.
This may result from the large
elastic modulus, and the relative small surface energy of diamond (the dangling bonds in the normal atmosphere are passivized by
hydrogen or oxygen atoms). Thus, MD calculations for the model studied above, but with increased Young's modulus $E> 1 \ {\rm GPa}$,
gives so small friction that it cannot be detected within the noise level of the simulations. The Young's modulus of diamond
$E\approx 1000 \ {\rm GPa}$ makes the ratio $w_0 / E$ much smaller than we used above even if $w_0 \approx 1 \ {\rm J/m^2}$.
The large modulus of diamond also results in incommensurate arrangements of the atoms at the sliding interface (unless two single crystals
with aligned crystal orientations are used) so for clean smooth surfaces 
one expects negligible contribution from the internal regions of the contact area.
The (small) friction observed in practical applications must be due to contamination molecules (see above).

\vskip 0.3cm
{\bf Acknowledgments}
J. Wang would like to thank scholarship from China Scholarship Council (CSC) and funding by National 
Natural Science Foundation of China (NSFC): grant number U1604131.

% QQ

\end{document}